\documentclass[prx,aps,floatfix,twocolumn,preprintnumbers,superscriptaddress]{revtex4-1}
\usepackage{amsmath}
\usepackage{amsfonts}
\usepackage{amssymb}
\usepackage{epsfig}
\usepackage{graphicx}
\usepackage{color}

\begin{document}

\title{Environment-dependent swimming strategy of Magnetococcus marinus under magnetic field}
 
\author{Nicolas Waisbord}
\affiliation{Institut Lumi\`ere Mati\`ere, CNRS UMR 5306, 
Universit\'e Claude Bernard Lyon1, Universit\'e de Lyon, France}

\author{Christopher T. Lef\`evre}
\affiliation{CNRS/CEA/Aix-Marseille University, Biosciences and Biotechnologies Institute, 13108 Saint Paul lez Durance, France}

\author{Lyd\'eric Bocquet}
\affiliation{Laboratoire de Physique Statistique, Ecole Normale Supérieure, UMR CNRS 8535, 75005 Paris, France}

\author{Christophe Ybert}
\affiliation{Institut Lumi\`ere Mati\`ere, CNRS UMR 5306, 
Universit\'e Claude Bernard Lyon1, Universit\'e de Lyon, France}

\author{C\' ecile Cottin-Bizonne}
\affiliation{Institut Lumi\`ere Mati\`ere, CNRS UMR 5306, 
Universit\'e Claude Bernard Lyon1, Universit\'e de Lyon, France}

\begin{abstract}

Magnetotactic bacteria (MTB) are fascinating micro-organisms which possess embodied biomineralized nanomagnets providing them  the ability to orient with the Earth's magnetic field. This property is presumably related to an evolutionary advantage in finding the oxic-anoxic interface along the up and down direction in aquatic environments. So far the magnetic field response by MTB,  called magnetotaxis, has been well described by a paramagnetic model where bacteria orient passively along the field lines according to a purely physical mechanism where magnetic torque and orientational Brownian noise compete.
Here we demonstrate using \textit{Magnetococcus marinus} strain MC-1 as MTB model that magnetotaxis shows more complex behaviors, which are affected by environmental conditions of different types. Indeed while MC-1 swimmers are found to essentially obey the paramagnetic paradigm when swimming in their growth medium, they exhibit a run-and-tumble dynamics in a medium devoid of energy source. Tumbling events are found to provide isotropic reorientation capabilities causing the cells to escape from their prescribed field direction. This behavior has a major influence on the capabilities of the cells to explore their environment across field lines and represents an alternative search strategy to the back-and-forth motion along field-imposed tracks.
Moreover, we show that aside chemical conditions, steric/geometrical constraints are also able to trigger tumbling events through obstacle encountering. Overall, physico-chemical environmental conditions appear to be important parameters involved in the swimming properties of MTB.
Depending on environmental conditions, the run-and-tumble mobility may provide advantages in the search for nutrient or ecological niche, in complement to classical magnetotaxis.

\end{abstract}

\maketitle


{M}any micro-organisms have the ability to move in their environment in order to respond to their needs, looking for instance for nutrients or for an optimum of oxygen concentration \cite{Wadhams:2004kq}. Among these micro-organisms magneto-aerotactic bacteria (MTB) are aquatic prokaryotes that show the specificity of a magnetically-assisted aerotaxis: they passively orient with magnetic field lines along which they actively swim, the so-called magneto-aerotaxis \cite{Frankel:1997fx,Blakemore:1975vq}. Combined with the fact that --except at the equator-- magnetic field lines are indicative of the top to bottom location, this constitutes the basis of the magneto-aerotaxis paradigm. Indeed,  by switching their motions from 3-dimensional to 1-dimensional (1D), MTB find more efficiently their ecological niche, that is  anaerobic or micro-aerobic conditions  \cite{Frankel:2009gd}. Thus, in the Northern Hemisphere, the local magnetic field is antiparallel to the oxygen gradient and MTB are north-seeking while in the Southern Hemisphere, the local magnetic field is parallel to the oxygen gradient and MTB are south-seeking.
The 1D motility of the cells is regulated back and forth to remain in the vicinity of the oxic-anoxic interface.

Magneto-aerotaxis is a common feature shared by a large number of bacterial species disseminated all around the globe in almost each and every freshwater and marine environment \cite{Lefevre:2013js}. MTB have in common the biomineralization of magnetosomes, a prokaryotic organelle, generally aligned in the cell and responsible for their magnetic orientation \cite{Blakemore:1975vq}. Our study is focused on the model magnetotactic bacterium \textit{Magnetococcus marinus} strain MC-1\cite{{Bazylinski:2013ju}}. This bacterium has a polar magneto-aerotaxis: in a homogenous chemical environment it swims towards the magnetic north (or south) for north-seekers in the Northern Hemisphere (respectively south-seekers) \cite{Frankel:1997fx, Lefevre:2014fs}. Since the early studies on the motility of MTB, the response to magnetic field alone has been well described by a paramagnetic model. In this framework, the bacterium direction results from a competition between (i) a passive orientation of the bacterium along the magnetic field lines associated to the torque exerted by the field on the magnetic moment of the bacterium and (ii) thermal energy that tends to disorient the bacterium \cite{Frankel:1984gv}. Within this description, MTB have no active control on their swimming orientation apart from back and forth switches.

Magneto-aerotaxis could be seen as a limitation in the ability  of MTB for optimizing their chemical environment and thus one may wonder if MTB can override this paramagnetic swimming behavior to let emerge other swimming strategies \cite{Simmons:2006ib}. This interrogation is legitimated by the fact that i) MTB are found around the equator \cite{Frankel:1981br} where the field lines no longer provide a relevant direction for exploring the oxygen gradient and ii) south-seeking MTB are sometimes found in the Northen Hemisphere \cite{Simmons:2006vv} suggesting that alternate search mechanisms may be possible.
In this contribution we explore the behavior of  strain MC-1 under various environmental and magnetic field conditions. Starting with a canonical experiment of chemical waves and bacterial aerotactic band, we first show that the field-mediated 1D motion strategy encompassed in the classical paramagnetic model cannot always account for the observed response of MC-1. Turning to model situations of dilute bacteria in homogeneous chemical environments, we evidence the existence of a swimming strategy alternative to the paramagnetic response. This strategy takes the form of a run-and-tumble swimming pattern by analogy to the well-known behavior exemplified by \textit{Escherichia coli} and more generally by a large variety of chemotactic bacteria. This strategy, which can override the magnetically-set orientation, is found to be promoted by the chemical composition of the surrounding environment but also by the geometry of the surrounding environment  through collisions with obstacles. 

Overall, our results provide a clear evidence that some MTB do not rely on the sole magnetically-assisted aerotaxis while swimming toward the North or South but are also capable of supplementary strategies that deflect from their prescribed field direction.

%
%
%
\section{Chemical waves under magnetic field}
%
%
%
%
\begin{figure}
\centering
\includegraphics[width=.4\textwidth]{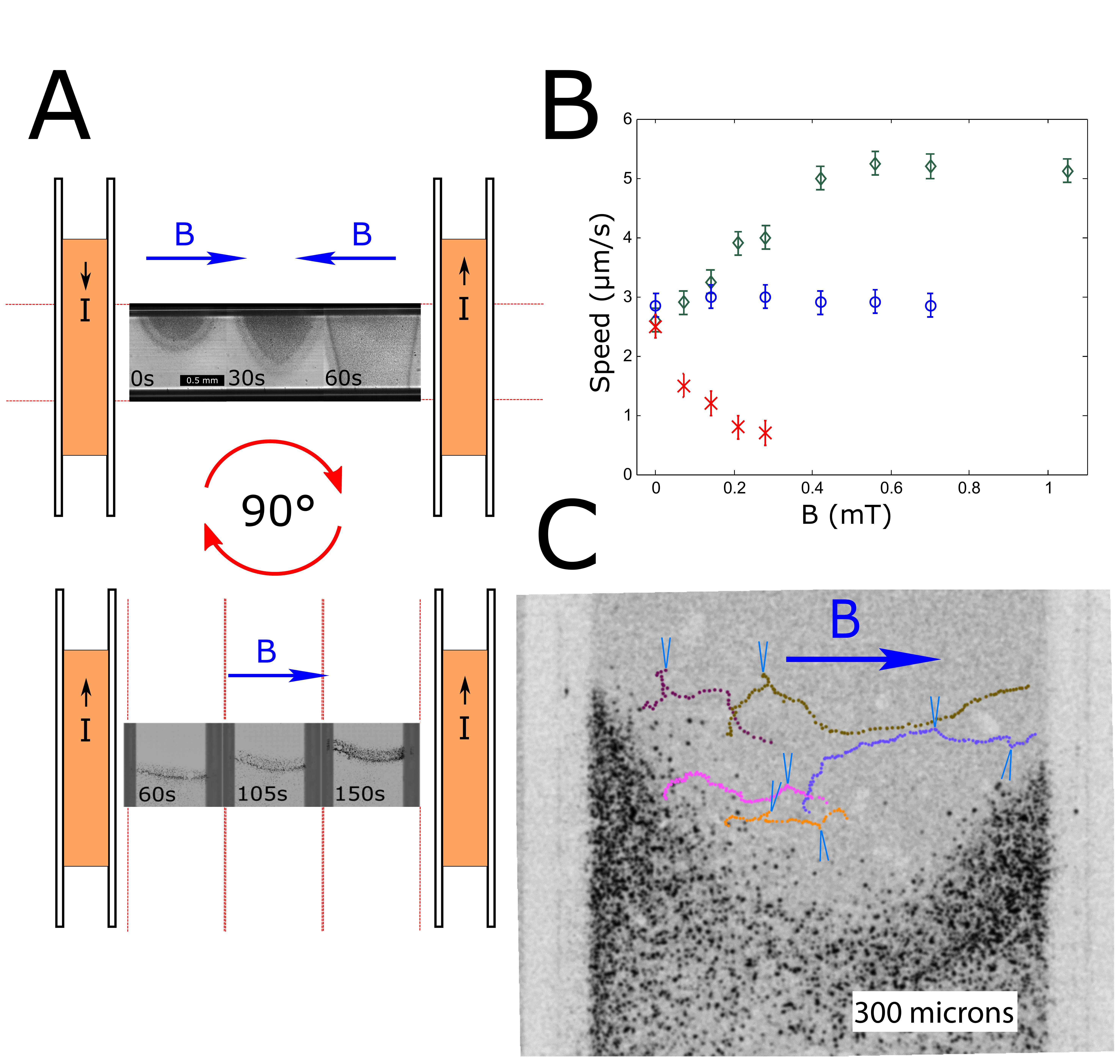}
\caption{\label{fig:Chemowave} A: Set up of the chemical wave experiment. Up: initial stage, the glass capillary aligned on the axis of a Helmoltz Coils set-up and MC-1 bacteria are concentrated in a central ``droplet'' using focusing magnetic fields from each Coil. After a few minute, a front grows from the droplet and a propagating chemical wave develops along the capillary. Down: 1D propagation stage, the magnetic field is set homogeneous by reversing one of the coils' current and the capillary is either oriented perpendicular or along (not shown) the field direction. Remarkably, bacteria band keep moving driven by the chemical gradient even when the magnetic field is perpendicular to the propagation direction. B: Speed of the chemical wave band with the magnetic field intensity for: perpendicular field  (blue $\circ$), and parallel field towards propagation (green $ \diamondsuit$) or against it (red $\times$). C: Example of tumbling tracks in the propagating front region, showing distinct tumbling events.}
\end{figure}

We performed the benchmark aerotactic band experiment commonly used to characterize the magneto-aerotactic behavior of MTB (axial, (di)polar,  or unipolar \cite{Lefevre:2014fs}). Unlike classical experiments however, where the magnetic field is pointing in the opposite direction of the oxygen gradients, we explored different field configurations. In the present experiments, MC-1 cells are introduced in a glass capillary filled with medium-rich solution (see Materials and Methods) that is subsequently sealed. Prior to introduction in the capillary, North seeking cells were selected using a magnet. Thus placing the capillary axis along the converging field direction obtained from Helmholtz coils leads to the formation of a high concentration of bacteria "droplet" at the center of the capillary (see figure \ref{fig:Chemowave} A). Few minutes after,  a band of bacteria that spontaneously propagates at a velocity around 3$\mu m/s$ is formed. Due to the high concentration of bacteria in the initial droplet and in the aerotactic band, chemical gradients develop and give rise to a propagation of the band (figure \ref{fig:Chemowave} A) as observed and described for other systems \cite{Frankel:1997fx, Saragosti:2011jk, Bennet:2014de}.

Once the band is formed, it is possible to rotate the capillary axis and to use the Helmholtz coils to see how  the band propagation responds to a uniform magnetic field. Most strikingly, we have measured the band velocity for magnetic fields perpendicular to the propagation direction (the capillary axis). As shown in figure \ref{fig:Chemowave} B, the remarkable observation is that the front propagation (i) persists and (ii) exhibits a velocity which is independent of the field magnitude.
This observation is different from what was previously observed for \textit{Magnetospirillum gryphiswaldense} strain MSR-1 \cite{Bennet:2014de} where the tilt of 90$^{\circ}$ of the magnetic field drastically reduced the magneto-aerotaxis efficiency. This difference likely comes from the different magnetotactic behaviors displayed by these two strains. Indeed, strain MSR-1 has an axial magnetotactic behavior where aerotaxis is prioritized over magnetotaxis, while strain MC-1 is a dipolar MTB where the cells give priority to following the direction indicated by the magnetic field.

Very clearly, such a motion associated to magneto-chemotaxis cannot fit into the existing framework of chemically-activated back and forth motion along the magnetic field lines. Note that few experiments have also been conducted where the field (i) was set homogeneous and parallel to the band propagation direction, or (ii) was stopped. In all cases, band propagation persisted with velocities not much affected. For instance figure \ref{fig:Chemowave} B shows the propagation speed along or against the magnetic field direction. While the field magnitude indeed either accelerates or decelerates the band, the field-assited contribution never exceeds the propagation velocity found for perpendicular field.

Somehow, such an observation is coherent with the fact that magnetotactic cocci related to strain MC-1 can be found in most aquatic environments on Earth, including at the equator, where the magnetic field is horizontal and thus perpendicular to the oxygen concentration gradient in water \cite{Frankel:1981br}. However, it requires extending the magneto-aerotactic paradigm to incorporate alternative swimming strategies which could overcome the constraint of the field direction. To collect clues of the underlying mechanism responsible for the cross field-direction propagation, figure  \ref{fig:Chemowave} C provides a zoom on the front region, where a few individual bacteria trajectories have been superimposed in colors. As marked by the arrow symbols along trajectories, it is possible to distinguish very clearly events of dramatic reorientation --showing up as kinks-- which allow the bacteria to swim across field lines. These events are reminiscent of a well-known chemotactic strategy of run-and-tumble performed by numerous bacteria among which \textit{E. coli}. Such run-and-tumble motility was observed in \textit{Magnetospirillum magnetotacticum} strain AMB-1 swimming in homogeneous chemical conditions \cite{Zhu:2014iva}.
In the following, we perform complementary experiments to verify this hypothesis of a  run-and-tumble strategy which can take over classical magneto-aerotaxis depending on environmental conditions.

%
%
%
%
\section{Chemically-induced tumbling of Magnetococcus marinus}
%
%
%
%
%
\begin{figure}
\centering
\includegraphics[width=.45\textwidth]{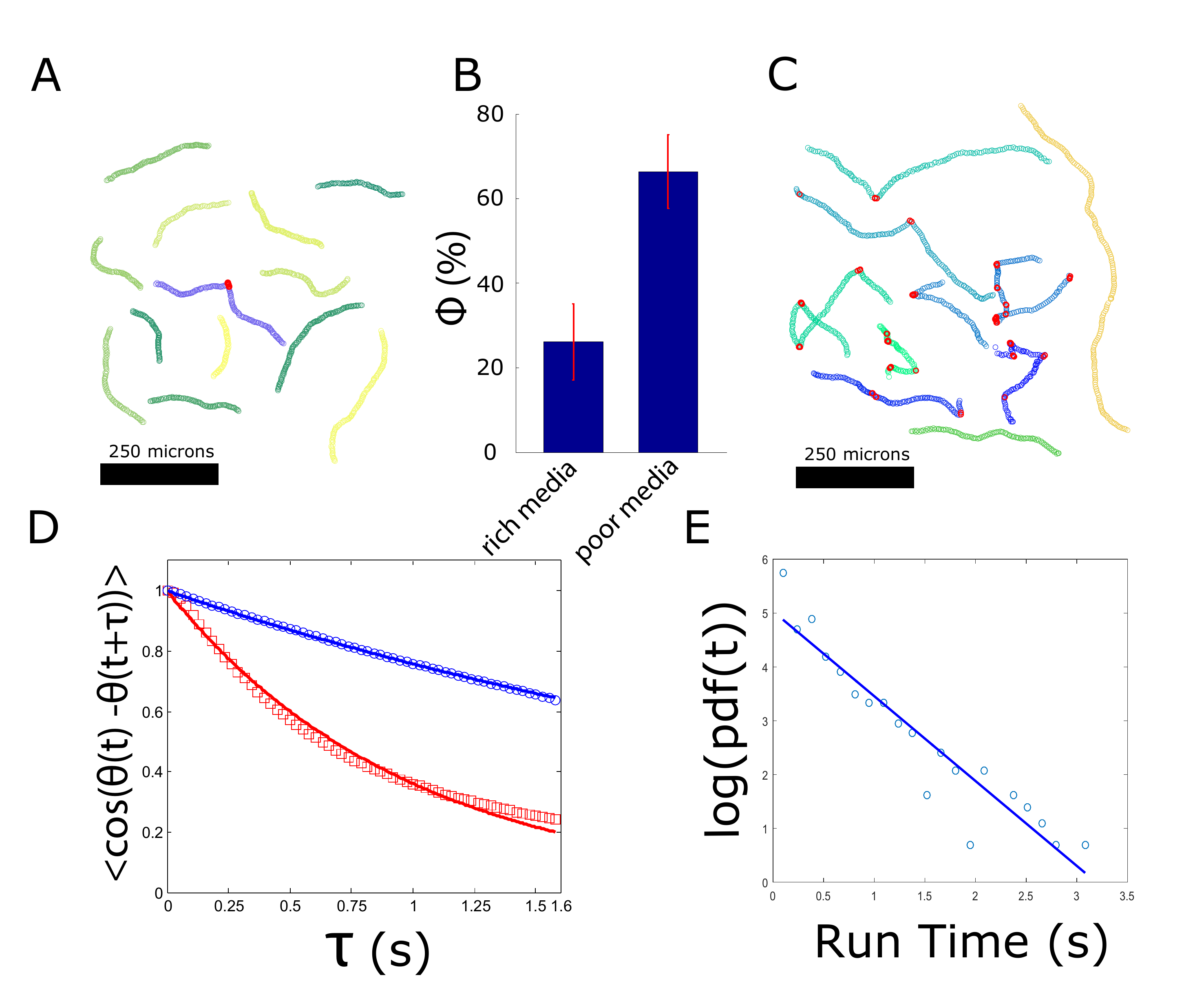}
\caption{\label{fig:Homogeneous conditions} A: Tracks in rich media without magnetic field: few tumbles (in red) are observed B: Percentage of tumblers for rich and poor media. C: Tracks in poor media without magnetic field: more tumbles (in red) are observed. D: Correlation function of the orientation for tumblers (red $\square$) and Smooth Swimmers (blue $\circ$) in the poor media without magnetic field. E: log of the probability density function of the running time between the detected Tumbles. The fit gives a running time of  0.6 $\pm$ 0.2 seconds}
\end{figure}

In order to get some insight into the MC-1 swimming strategy, and to characterize the proposed run-and-tumble mechanism, we used dilute suspensions of bacteria introduced into a PDMS (polydométhylsiloxane) channel (see Materials and Methods) that had been filled with homogeneous medium, either denoted as rich (containing thiosulfate, the energy source used by MC-1) or poor (without thiosulfate). Note that this poor medium appears closer to environmental conditions.
Under these experimental configurations, single individuals dynamics could be followed under magnetic field, with no evolution of the properties over 20 min.

\subsection{No field}

We start by examining the swimming behavior in the absence of --added-- magnetic field. Note that the earth magnetic field at our latitude is around 17$\mu$T, for which our cultured strains showed no biased directional motion. Figure \ref{fig:Homogeneous conditions} A shows a typical set of individuals tracks captured for MC-1 in the rich medium. Each track has been analyzed in search for tumbling events, here defined as locations along the tracks which combine high angular and low translational velocities (\cite{Masson:2012ky} and Materials and Methods). The detected events are indicated as red positions on the experimental tracks. In the rich medium, most trajectories were found to be smooth with only a minority containing tumbling events as can be seen in Figure \ref{fig:Homogeneous conditions} A. On the contrary, when considering tracks recorded in the poor (thiosulfate-free) medium (figure \ref{fig:Homogeneous conditions} C), tumbling trajectories were found to be dominant with only a minority of remaining smooth trajectories. More quantitatively, figure \ref{fig:Homogeneous conditions} B gathers the overall percentage of trajectories showing tumbling events in the two different media. The amount of tumbling trajectories evolves markedly from $26\pm 9\%$ in a homogeneous rich medium up to $67\pm 9\%$ in the poor medium (calculated respectively over 1000 and 1500 trajectories). This is consistent with our observation for chemical waves under perpendicularly oriented magnetic field, \textit{i.e.} strong reorientation events --tumbling-- can be promoted by the chemical environment.

Concentrating on either smooth or tumbling trajectories alone, we have characterized the auto-correlation function of the swimming orientation. Figure \ref{fig:Homogeneous conditions} D shows the normalized experimental correlation functions together with their adjustment with an exponential decay $\langle\cos\theta(t)\cos\theta(t+\tau)\rangle=\exp(-t/\tau)$. The effect of the tumbling events on the persistence time of the swimming orientation is evident, with decorrelation times of, respectively, $\tau= 3.6\pm 0.4\,$s for smooth trajectories and $\tau= 1.0\pm 0.2\,$s for tumbling trajectories. More precisely in the case of the so-called smooth trajectories, we expect the dynamics to correspond to persistent random walk \cite{Berg:1972bq,Berg1993}. Within this configuration, we expect the decorrelation time for orientation to read $\tau=\tau_R/2$ with $\tau_R$ the random noise orientation time scale of expression $8\pi\eta R^3/kT$, with $\eta$ the surrounding viscosity, $k$ the Boltzmann constant, $T$ the temperature and $R$ the bacterium radius. With a typical radius around $1\,\mu$m \cite{Meldrum:1993id,Bazylinski:2013ju} this would predict a decorrelation around 3.0\,s in perfect agreement with the measured value. Now turning to the tumbling trajectories, beyond the decrease of the correlation time already reported, it is possible to look at the distribution of running times, which separate each tumbling event. As shown in figure \ref{fig:Homogeneous conditions} E, the measured density probability function  is found to be exponential, with a characteristic decay time of $0.6\pm 0.2$\,s in agreement with the orientation decorrelation time. This is reminiscent of the characteristic Poisson distribution of the run times as found for \textit{E.coli}  \cite{Berg:1972bq}, and further legitimates the suggested analogy with run-and-tumble trajectories. In some aspects, it echoes recent work on another magnetotactic bacterium, \textit{Magnetospirillum magneticum} strain AMB-1, which showed that the response {to the magnetic field} of MTB can incorporate active pathways \cite{Zhu:2014iv}. Here we do also identify an active response into the magneto-aerotactic bacteria dynamics; however not related to the magnetic field action but to the chemical environment.

\begin{figure}
\centering
\includegraphics[width=.45\textwidth]{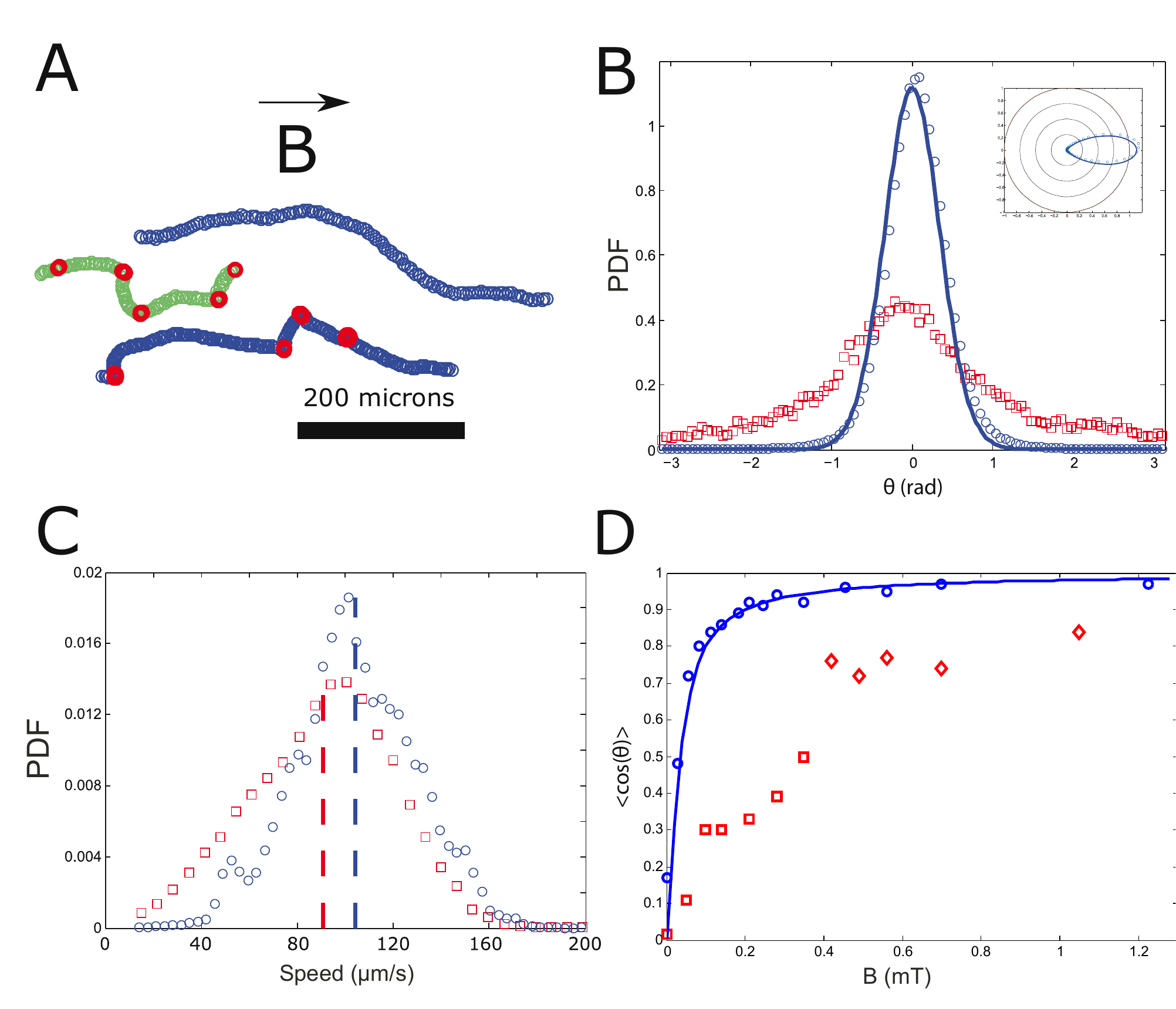}
\caption{\label{fig:Magnetic field} A: Tracks under magnetic field magnetic field (poor medium). B: PDF $p(\theta)$ of the swimmers orientation under field ($B=0.21\,$mT) for smooth (blue $\circ$) and tumbling (red $\square$) swimmers. Inset: polar diagram for smooth swimmers fitted by paramagnetic Langevin model. The fit for the smooth swimmers by  $p(\theta) \propto exp(MBcos(\theta)/kT)$ gives $M=1.7 \times 10^{-16} A.m^2$  C: Same as in B, for the PDF of velocity magnitude; dashed lines corresponds to the mean velocities (respectively $91\pm25\,\mu$m/s and $104\,\mu$m/s). D: Normalized averaged axial velocity as a function of the magnetic field for smooth (blue $\circ$) and tumbling (red $\square$) swimmers. The blue line is the fit of this orientation of the smooth  swimmers by the Langevin function coth$\left(\frac{MB}{kT}\right)- \frac{kT}{MB} $.}
\end{figure}

\subsection{Under homogeneous magnetic field} 

\ Under magnetic field, as shown in figure \ref{fig:Magnetic field} A, cell's trajectories continue to exhibit two distinct behaviors: smooth and run-and-tumble. 
For the smooth swimmers, in figure \ref{fig:Magnetic field} B we see the polar orientation diagram (inset) together with the probability distribution function (PDF) of orientations. Starting from an isotropic distribution without field, the swimming direction gets peaked around the field direction, with increasing directionality as the magnetic field increases. Under the classical description, we expect the bacteria orientations to be passively set by the magnetic torque exerted on magnetosomes, with the finite width of the distribution arising from the orientation noise attributed to rotational Brownian motion. Indeed the figure \ref{fig:Magnetic field} B shows that the PDF of smooth swimmers is perfectly fitted by a Langevin paramagnetic prescription of the form PDF$\propto\exp([MB/kT]\cos\theta)$. Moreover, the fitted distribution yields the value of the reduced energy $MB/kT= 8 \pm1$. Assuming  that the only source of orientational noise comes from thermal Brownian motion, this provides a value of the magnetic momentum carried by each bacterium of $M\approx1.7\pm0.2\,10^{-16}$\,S.I., consistent with direct measurements of the momentum (see Materials and Methods). Accordingly, the average axial velocity (along the field direction) is perfectly fitted by the prediction for paramagnetic orientating swimmers as can be seen in figure \ref{fig:Magnetic field} D. Note that the magnetic field dependency was also checked and found to agree quantitatively with the paramagnetic approach \cite{Blakemore:1975vq}. Therefore, the smooth swimmers, which dominate the observed dynamics in rich medium, are  corresponding to the previously reported response of MTB under field. It further ensures the coherence of our observations with previous reports \cite{Nadkarni:2013dk}.

Concerning the run-and-tumble dynamics, we first compare the velocity PDFs of smooth and tumbling swimmers in figure  \ref{fig:Magnetic field} C. As can be seen, the swimming activity appears unmodified in tumbling conditions, with a mean velocity essentially unchanged around $100\pm25\,\mu$m/s and marginal changes in the distribution. Indeed, the tumbling PDF differs only by a little skew towards low velocities which is coherent with the fact that tumbling events are associated to locally high rotational velocities and low translational ones.
An ``active magnetotaxis'' model was proposed for Magnetospirillum magneticum strain AMB-1, where the magnetotactic cells would sense the magnetic field thanks to a chemotaxis protein that would interact with the magnetosome chain and transfer the magnetic signal to the flagella \cite{Philippe:2010da}. 
We have measured the tumbling rate as a function of the magnetic field and have seen no dependency which leaves the question of the challenging model of ``active magentotaxie" not consistent with our observation in strain MC-1.

\begin{figure}
	\centering
		\includegraphics[width=0.4\textwidth]{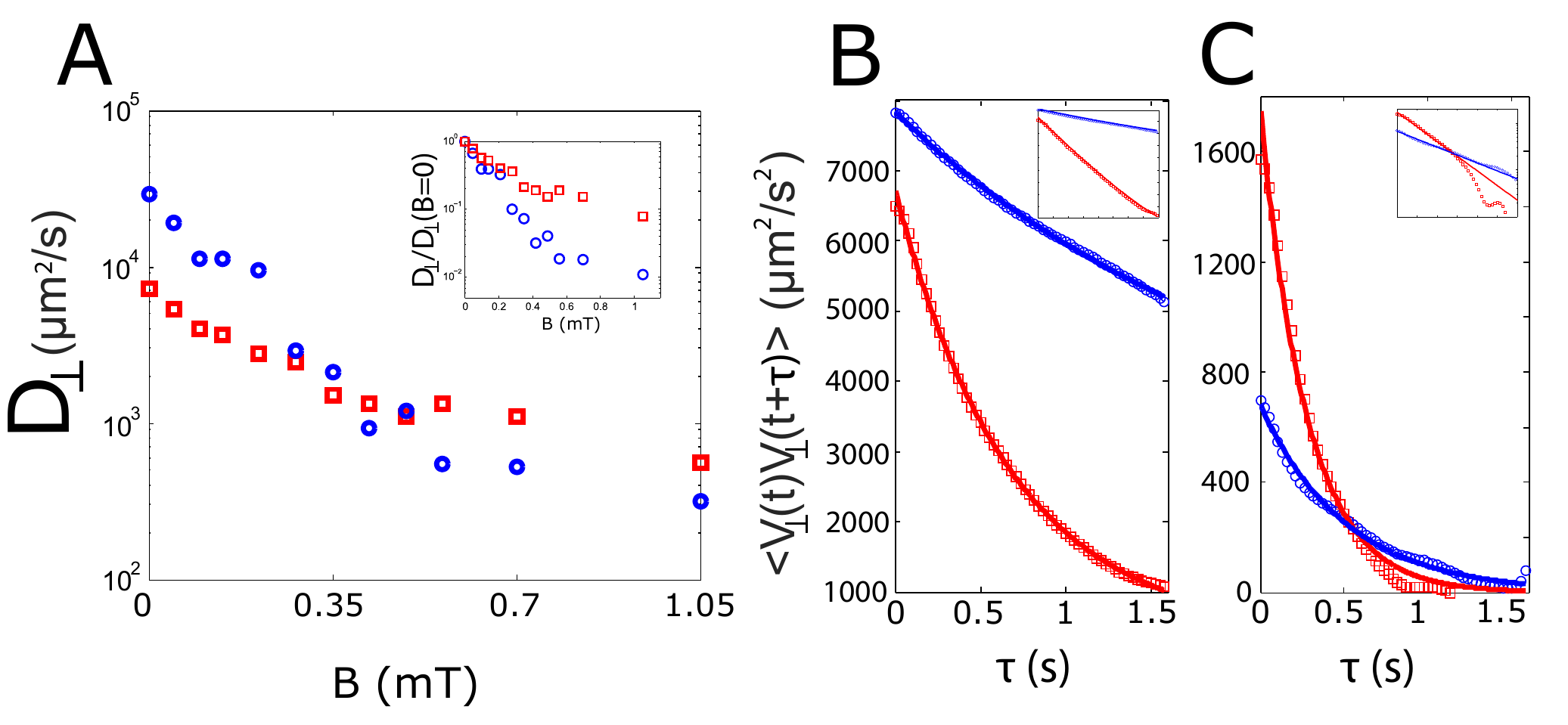}
	\caption{\label{fig:Diff}A: Perpendicular diffusion coefficient for smooth (blue $\circ$) and tumbling (red $\square$) swimmers as a function of the magnetic field. B: Same as in A for correlation function of perpendicular velocities under no field. C: Same as in B under 0.7mT magnetic field.}
	\end{figure}

The ability of bacteria to escape the direction set by the external magnetic field can be  quantified through the diffusion coefficient $D_\perp$ in the direction perpendicular to the imposed field. In practice, this is measured from the correlation function according to the Green-Kubo relationship: $D_\perp=\int_0^\infty \langle v_\perp(t)v_\perp(0)\rangle\,dt$. Figure \ref{fig:Diff}  shows the cross-diffusivity for smooth and tumbling swimmers as a function of the external field, the two appearing markedly different. Under low fields, smooth swimmers explore unexpectedly more space around the magnetic field line than tumblers. Indeed the bacteria have an effective diffusion coefficient that is proportional to $V^2\tau_r$ where $\tau_r$ is the rotational diffusion time \cite{Palacci:2010hk}. Tumbling can be seen as a reduction in $\tau_r$ which implies a reduction in cross-direction explored and thus in the effective diffusion coefficient $D_\perp$ (figure~\ref{fig:Diff} B). However, the situation gets reversed at higher fields (figure~\ref{fig:Diff} C) where tumbling swimmers become the most efficient at exploring the environment. By allowing initial conditions spanning the all directions, tumbling events allow excursions in the transverse direction which are statistically hardly possible for the smooth trajectories that get more and more confined along the field direction.  The range of spatial exploration by the tumbling swimmers does not depend on the applied magnetic field (inset figure~\ref{fig:Diff} A)

Overall, we have evidenced that depending on the chemical environment, MC-1 can show very different swimming behaviors. When studied in the rich medium, bacteria display smooth trajectories oriented on average by the magnetic field. These trajectories are illustrative of persistent random walks in external fields, which combine (i) constant velocity swimming, (ii) random noise and (iii) external torques which compete for the swimmers' orientation. Indeed, the classical paradigm for passive magnetotaxis is well recovered under these circumstances, even quantitatively. However, as already pointed out, growth media used in the lab are generally much more concentrated in energy source than what bacteria usually found in their environment \cite{Stocker:2012ea}.

While most studies that sought to decipher the magnetotactic behavior analize bacteria in their growth conditions \cite{Lefevre:2014fs}, here we show that the use of a poor medium --arguably closer to environmental conditions-- conduct to the switch into a run-and-tumble dominated swimming. This newly discovered strategy has a deep impact on the cross-direction diffusivity of bacteria and could thus be an advantage in poor conditions for efficiently finding the ecological niche.

%
%
\section{Confinement-induced tumbling}
%
%
\begin{figure*}
\centering
\includegraphics[width=0.8\linewidth]{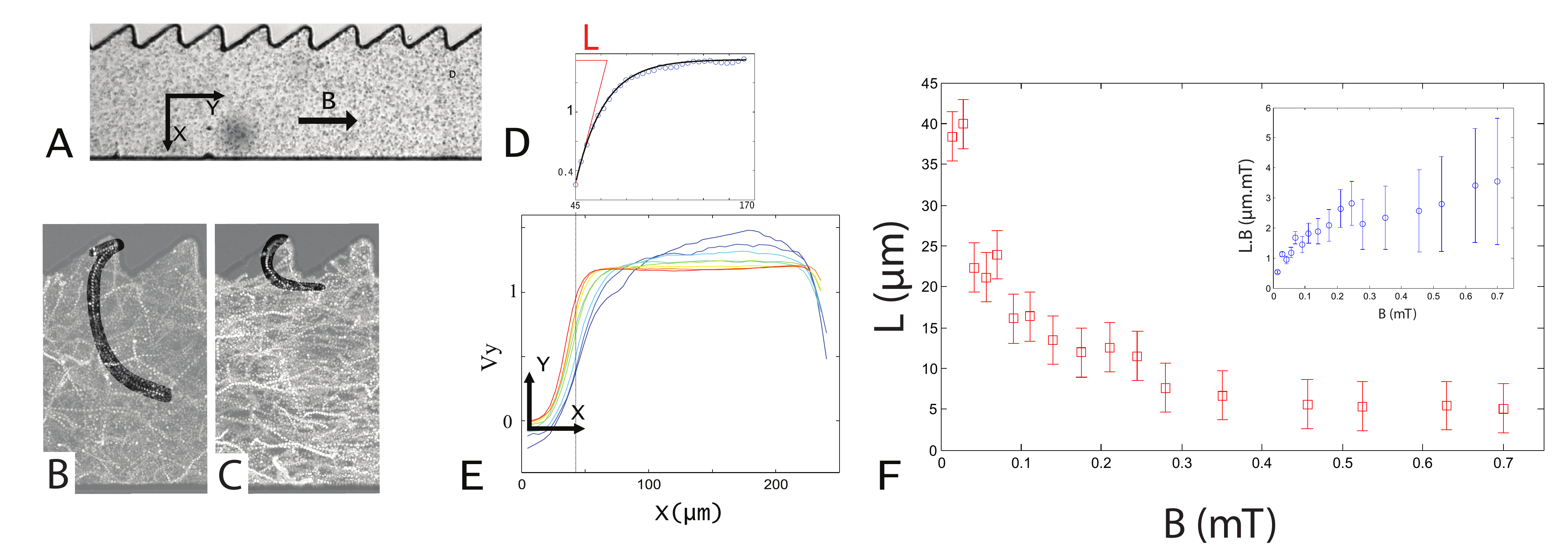}
\caption{\textbf{A}: Force-free rheological experiment: bacteria in structured micro-channels. \textbf{B}: Bacteria tumbles upon wall-collision, followed by trajectories during orientation relaxation  at $140\,\mu$T. \textbf{C}: Same as B for a field of $490\,\mu$T. \textbf{D}: Velocity Profile averaged alongside the channel (from tooth extremities up to flat saturated velocity); Solid line correspond to an exponential profile $A\exp(-x/L)$, thus yielding the wall influence length $L$. \textbf{E} Velocity profiles for various magnetic fields. \textbf{F}: Variations of $L$ under various magnetic field $B$. Inset:  $LB$ vs $B$.
}
\label{fig:fig5}
\end{figure*}

{Besides the chemical composition of the environment surrounding the MTB, we show that other environmental constraints can induce changes in their swimming behavior under the form of tumbling events. These constraints are of steric or geometrical type as can be encountered in natural environment such as sediments.}

Experimentally, these effects have been probed in a model geometry made up from a straight microfluidic channel which posseses ``rough'' structured walls. Dilute MC-1 cells  (volume fraction less than 0.1\%) are suspended in the rich medium in order to minimize the intrinsic amount of tumbling. Again, north-seekers have been selected upon filling with a magnet, and we hereafter consider all swimmers as smooth swimmers. Bacteria are then submitted to a homogeneous magnetic field along the channel axis and  their (swimming) motion inside the micro-channel  is analyzed (figure \ref{fig:fig5} A). Note that for comparison purposes, channel walls have been made asymmetric with one flat wall and one sawtooth textured wall. As evidenced in figure \ref{fig:fig5} B-C, a bacterium that collides with the wall will eventually reorient itself until it swims again away from the wall. This major reorientation in trajectories can again be assimilated to a tumbling event, now occuring in response to a mechanical constraint rather than a chemical one. Interestingly, recent studies have demonstrated close behaviors triggered by wall encounters. For instance \textit{Magnetospirillum magneticum} strain AMB-1 has been shown to switch from north-seeking swim to south-seeking swim when meeting an interface \cite{Reufer:2014kt} while cells of strain  MO-1  (a magnetotactic cocci related  to MC-1) was demonstrated to be able of making U-turns when encountering obstacles \cite{Zhang:2013cf}. 

We now look at the magnetic field influence on wall-tumblers. We observe that, when the field increases, the magnetic reorientation --that follows the tumbling event-- is more effective yielding to shorter re-injection length into the main bacteria stream (figure \ref{fig:fig5} B-C). Increasing the bacteria concentration up to $\approx 1\%$, we can run Particule Images Velocimetry algorithms to access the velocity profiles of bacteria within the microchannel. Figure \ref{fig:fig5} D shows the asymmetry of the velocity profiles obtained, which matches the asymmetry of the walls structuring. Indeed the effect of the walls is related to the previously called re-injection length --that is the average exploration length towards the channel center-- that the bacteria are able to travel before being reoriented. Experimentally, the length is defined from velocity profiles by fitting them with an exponential relaxation and defining the \textit{influence length $L$} as the decay length (figure \ref{fig:fig5} E). 

In Figure \ref{fig:fig5} F, we represent the evolution of the influence length $L$ associated to sawtooth-shaped walls on the velocity profile of the suspension. Starting from a finite length at vanishing magnetic field, this influence length decreases with increasing field. Plotting $LB$ rather than $L$ reveals the existence of a plateau value at high fields, thus suggesting that the influence length decreases as $1/B$ in this limit. Concentrating first on the small field limit, once a bacterium has tumbled at the wall and is re-injected into the main stream, the wall influence is bounded by the persistence orientation time in dilute-regime and so is the influence length $L\propto V_0\tau_R$.

In the high field limit, the relaxation time towards the field direction becomes shorter than the Brownian persistent time and thus dominates the extent to which wall influence propagates. The ratio between both times writes
\begin{equation}
\frac{\tau_R}{\tau_B} =  \frac{8\pi\eta R^3}{k_B.T}\left(\frac{8\pi\eta R^3}{M.B}\right)^{-1} = \frac{M}{k_B.T} \times B.
\end{equation}
According to the orientation distribution (Fig. \ref{fig:Magnetic field}), this ratio is around 4 for $B=0.1\,$mT, for which we do observe the plateau in $LB$ (figure \ref{fig:fig5} F). This confirms the domination of the magnetic alignment relaxation time in this regime, for which we then expect the $L\propto1/B$ dependency according to
\begin{equation}
L \propto V_0\tau_B=V_0\frac{8\pi\eta R^3}{M.B}.
\end{equation}

Overall, flow profiles close to sawtooth-shaped walls suggest that the effect of an obstacle on bacteria dynamics can be summed up by a wall tumbling followed by an orientation relaxation over a length scale either dominated by a persistent random walk exploration (low field) or a deterministic trajectory under magnetic field. In the environment, obstacles are many and the probability to meet a particle is high particularly in sediment where MC-1 and other MTB are generally found. Such tumbling behavior allows the cells to avoid obstacles and to continue swimming toward their preferred oxygen concentration.

\section{Conclusion.}

Our results show that the motility of  strain MC-1 is affected by the environment.
The magneto-aerotaxis model is not just a passive alignment of the cells while they are swimming. Indeed, whereas the cells are constrained to swim in one direction due to the magnetic forces they are still able to sense their environment and to swim in other directions if the chemotactic machinery of the cells detects the need to do so.  In a rich medium bacteria exhibit a smooth motility perfectly described by a Langevin paramagnetic model. In a poor medium, closer to environmental conditions, we show  that MC-1 cells can exhibit a run-and-tumble motion outstripping the simple paramagnetic behavior.  By changing their direction of motility in poor media bacteria increase the chance to meet their energy source while swimming around the oxic-anoxic interface. 
Furthermore we show that that tumbling can also occur under geometrical constrains as encountered in crowded environments of sediments for instance.

Finally, in a wider perspective, let us note that when under the conditions for a paramagnetic swimming response, magnetotactic bacteria form a remarkable active system of extremely efficient self-propeled individuals whose orientation can be quantitatively controlled by a physical magnetic field \cite{Bazylinski:2004jx}. In the blooming context of biological active matter or in the perspective of drivable in-body micro-swimmers for medical applications, this system forms a very appealing model of controllable active system.

\section{Materials}
\textbf{Bacterial strain.} 
Cells of \textit{Magnetococcus marinus} strain MC-1 were grown in a slightly modified semi-solid medium described in \cite{Bazylinski:2013ju}.The medium consisted of an artificial seawater (ASW) containing (per litre): NaCl, 16.43 g; MgCl$_2$.6H$_2$O, 3.49 g; Na$_2$SO$_4$, 2.74 g; KCl, 0.465 g; and CaCl$_2$.2H$_2$O, 0.386 g. To this was added (per litre) the following prior to autoclaving: 5 mL modified Wolfe's mineral elixir \cite{Frankel:1997fx}, 0.25 g NH$_4$Cl, 2.4 g HEPES, 100 $\mu$L0.2\% (w/v) aqueous resazurin and 2.0 g agar noble (Difco). The medium was then adjusted to pH 7.0, boiled to dissolve the agar and autoclaved. After the medium had cooled to about 45°C, the following solutions were added (per liter) from stock solutions (except for the cysteine, which was made fresh and filter-sterilized directly into the medium): 2.8 mL 0.5 M potassium phosphate buffer, pH 6.9; L-cysteine, to give a final concentration of 0.4 g.L$^{-1}$; 0.5 mL vitamin solution \cite{Frankel:1997fx}; 3 mL 0.01 M FeSO$_4$ dissolved in 0.2~M HCl; 3 mL 40\% sodium thiosulfate pentahydrate solution (i.e., 10mM of energy source); and 2.7 mL 0.8 M NaHCO$_3$ (autoclaved dry; sterile water added after autoclaving to make the fresh stock solution). The medium (10 mL) was dispensed into sterile, 15 x 125 mm screw-capped test tubes. All cultures were incubated at 28$^{\circ}$C and, after approximately 1 week, a microaerobic band of bacteria formed at the oxic-anoxic interface (pink-colorless interface) of the tubes. This band contains a majority of north-seeking cells; the rare south-seeking cells were eliminated via a magnetic separation that consists in transferring the cells from their semi-solid medium to a liquid ``swimming medium''.

\vspace{0.2cm}		
\textbf{Swimming media.}
The medium referred as \textit{rich medium} consisted of ASW to which was added (per litre) the following: 5 mL modified Wolfe's mineral elixir, 0.25 g NH$_4$Cl, 2.4 g HEPES. The medium referred as \textit{poor medium} consisted of the same medium as the rich one devoid of thiosulfate. Both media have a pH adjusted to 7. 

\vspace{0.2cm}	
\textbf{Microfabrication.}
Microchannels were fabricated by standard ``soft lithography'' technique: a mold made of negative photoresist (SU8 3100 Microchem) is obtained by conventional photolithography. Channels are then made of PDMS poured on these molds, at 70$\,^{\circ}\mathrm{C}$ for 3 hours, with a typical thickness of 1mm.
After cross-linking, the PDMS imprint is pealed of its mold. The channels are cut, the entrance and the outlet are punched, and are finally bonded with a glass slide after an air plasma treatment. The obtained channels are then filled with swimming media by capillarity. The outlet is then sealed with capillary wax. A bacteria drop is then taken from its growing medium and then guided inside the channels with a magnet. Once bacteria are inside, the entrance is sealed with the wax.

\vspace{0.2cm}	
\textbf{Coils control.}
Coils are controlled directly from the computer, with a Labjack U12 device and an homemade current source. They can be set in different configurations, generating Magnetic field parallel or anti-parallel. We use the anti-parallel configuration to concentrate bacteria, and at the same time sort the south seekers from the north seekers. The parallel configuration allows fields ranging from 0 to 2.8mT, considered homogeneous in the observation zone of 1.2~mm. Alignment of channels with the coils is made  as follows:  coils are designed with a rectangular hole where a Nikon Mir Glass Slide fits exactly: the field and the line of the mir are perpendicular. The camera observation is then aligned on the mir. Then the micro channel is aligned with a ROI crop selection on the screen on the acquisition system.

\vspace{0.2cm}	
\textbf{Preparation of the experiment.}
We used linear silicone polydimethylsiloxane (PDMS) micro-channels (1200$\mu$m $\times$ 70$\mu$m $\times$ 5mm). The channels were filled with the swimming medium by capillary suction, then sealed on one side. A drop of a culture of strain MC-1 was disposed at the extremity of the channel and steered inside the micro-channel with a simple stirring magnet. Once the bacteria inside the channel, this latter is sealed on the former entrance. Using a computer controlled Helmholtz coils system we control the magnetic field inside the channel. We concentrate the north seeking bacteria inside the channel by using a configuration where coils generate fields that face one another and then we set the magnetic field to zero.

\vspace{0.2cm}	
\textbf{Microscopy and data acquisition.}
To acquire images we use a Leica DMI 4000B Microscope in transmitted light with a Zeiss 5X long working distance objective, equipped with a Baumer HXC40 4Mpx CCD camera controlled by custom Labview codes, covering a 1200$\mu$m$\times$ 1200~$\mu$m area. We run both rheology and single particle experiments at 35 fps. The long-working distance objective allows a depth of field of $\approx$100 $\mu$m, making possible to follow bacteria and acquire the projection of their trajectories. In single particle experiment, we assume that the problem is axially symmetric, which allows several calculus tricks for the relationships between projected observables and 3D ones.

\vspace{0.2cm}	
\textbf{Tracking and trajectory analysis.}
Image treatment and tracking was performed using standard tracking routine written with Matlab software (Mathworks). The selection of the trajectories was based on duration (>1.5 s) exploration radius (> 50$\mu$m) and average instant speed (> 40$\mu$m.s$^{-1}$).  To avoid taking into account the Helical Motion of the bacteria, we smoothed the positions by a gaussian filter on a width of 4 positions, so that the projected orientation of motion we consider in this study is the one of the main component of the motion. The projected observed speed is assumed to be the real one. The Transverse projected MSD is assumed to be half of the real one. The correlation function of the projected motion is assumed to be the same as the one in 3D. 
Finally, the projection of the Orientation Distribution function are fitted by the $\frac{I_1(\xi.cos(\alpha))+L_{-1}(\xi.cos(\alpha))}{4\xi^{-1}\sinh(\xi)}$, which is the statistics of the orientation expected of the Langevin Distribution, projected in the focal plane. 
Motivated by the large persistence lengths of the trajectories (over 200$\mu m$ for smooth swimmers), we computed $D_\bot(B)$, the diffusion coefficient perpendicular to the magnetic field, from the Green-Kubo formula, which reads:
\begin{equation}
2D_\bot(B) = \int_0^\infty \langle V_\bot(\tau) V_\bot(\tau+t) \rangle_\tau dt.
\end{equation}
We obtain $D_\bot(B)$ from the correlation functions of the speed in the transverse directions, which necessitates to follow bacteria during a shorter period of time and can be fitted with exponential functions both for the tumblers and the smooth swimmers.

\vspace{0.2cm}	
\textbf{Tumble detection.}
Inspired from the method of \cite{Masson:2012ky}, we define a function which combines the variation of the speed of the bacteria and their rotational acceleration, both characteristic of tumbling event where a bacteria stops and turns:

$f_{\alpha,\beta}(t) = \frac{|\omega(t)|^\alpha}{|V(t)|^\beta}$

For a smooth swimmer, this quantity cannot increase a lot more than 4 times its median value computed along a trajectory, as the disorientation is well described by a White Gaussian Noise. For a tumbler, the median reflects  its smooth swimming behavior as, for a clear majority of its time it is indeed smooth swimming, and the tumbling events are not visible in the computation of the median. These aredetected, and we compare it to 10 times the median of $f_{\alpha,\beta}(t)$ for each trajectory. This threshold and the exponents $(\alpha,\beta)$ are adjusted and for our data, a threshold of 10 and $(\alpha,\beta)=(2,2)$ are satisfactory. 

\vspace{0.2cm}	
\textbf{PIV analysis.}
PIV analysis was performed using the PIVlab OpenSource Matlab toolbox. Images had their background subtracted before PIV was run. The average velocity was done over 400 images, along 15 periods of the waved wall.  We have also shown, in a force free rhoelogy experiment, that tumbling at the wall can be induced by roughness which modifies the shape of the velocity profile.

\vspace{0.2cm}	
\textbf{Paramagnetic behavior in rich medium.}
An important physical parameter to estimate for a magnetic bacterium is its magnetic moment. For this we have carried out an experiment where non swimming bacteria are submitted to a gradient of magnetic field: living bacteria are suspended in a liquid rich medium with 0.5\% BSA  and buffered at pH 10. This high pH kills the bacteria while preserving the cell integrity and the BSA prevents them from sticking to the glass capillary in which the suspension is sucked. The capillary is then disposed in the middle of two facing magnets (Supermagnets W-03-N Gottmadingen, Germany) creating a magnetic gradient of $\left\|\vec{\nabla} \vec{B}\right\| = 80 T.m^{-1}$. Bacteria are dragged across the channel under the magnetic force created by such a gradient of field and the magnetic momentum is estimated by measuring the average velocities of the bacteria:

 $M = \frac{6 \pi \eta R.V_{drag}}{\left\|\vec{\nabla} \vec{B}\right\|}$.

We measure a magnetic momentum  $M=1.3  \times 10^{-16} A.m^{2}$  with a standard error $0.5 \times 10^{-16} A.m^{2}$. This value is obtained by an overage over 500 trajectories of bacteria coming from three different batches and it is within the range of magnetic momentum obtained for other magnetotactic bacteria~\cite{Nadkarni:2013dk}.

%
%
		
\vspace{0.5cm}

\begin{acknowledgments}
We thank Eleonora Secchi for her help with the data analysis.
We thank Veroniquie Utzinger, Guy Condemine and Nicole
Cotte-Pattat for their help with the bacterial cultures.
We thank INL for access to their clean room facilities
and AXA research fund for its financial support.  C.T.L. was supported by the French National Research Agency (GROMA: ANR-14-CE35-0018-01) 
\end{acknowledgments}

\end{document}